%% file: main.tex
\documentclass[sigconf, nonacm]{acmart} 
\pdfoutput=1
\newcommand\vldbdoi{XX.XX/XXX.XX}

\newcommand\vldbavailabilityurl{https://github.com/wiegerpunter/omnisketch}
\newcommand\vldbpagestyle{plain} 
\newcommand\sketch{\ensuremath{\mathbb{S}}}

\usepackage{algorithm}
\usepackage[noend]{algpseudocode}
\usepackage{subcaption}
\usepackage{cprotect}

\newcommand\todo[1]{\protect \textcolor{brown}{[TODO: #1]}}

\newcommand\simax{\ensuremath{S^{\max}_i(j, k)} }

\newcommand{\eat}[1]{}

\begin{document}
\title{OmniSketch: Efficient Multi-Dimensional High-Velocity Stream Analytics with Arbitrary Predicates}
  
\author{Wieger R. Punter}
\affiliation{%
  \institution{Eindhoven University of Technology}
  \streetaddress{Den Dolech 2}
  \city{Eindhoven}
  \country{The Netherlands}
  \postcode{5600MB}
}
\email{w.r.punter@tue.nl}

\author{Odysseas Papapetrou}
\orcid{n}
\affiliation{%
  \institution{Eindhoven University of Technology}
  \streetaddress{Den Dolech 2}
  \city{Eindhoven}
  \country{The Netherlands}
}
\email{o.papapetrou@tue.nl}

\author{Minos Garofalakis}
\orcid{}
\affiliation{%
  \institution{ATHENA Research Center \& Technical Univ. of Crete}
}
\email{minos@athenarc.gr}

\begin{abstract}
A key need in different disciplines is to perform analytics over fast-paced data streams, similar in nature to the traditional OLAP analytics in relational databases -- i.e., with filters and aggregates. Storing unbounded streams, however, is not a realistic, or desired approach due to the high storage requirements, and the delays introduced when storing massive data. Accordingly, many synopses/sketches have been proposed that can summarize the stream in small memory (usually sufficiently small to be stored in RAM), such that aggregate queries can be efficiently approximated, without storing the full stream. However, past synopses predominantly focus on summarizing single-attribute streams, and cannot handle filters and constraints on arbitrary subsets of multiple attributes efficiently.
In this work, we propose OmniSketch, the first sketch that scales to fast-paced and complex data streams (with many attributes), and supports aggregates with filters on multiple attributes, dynamically chosen at query time. The sketch offers probabilistic guarantees, a favorable space-accuracy tradeoff, and a worst-case logarithmic complexity for updating and for query execution. We demonstrate experimentally with both real and synthetic data that the sketch outperforms the state-of-the-art, and that it can approximate complex ad-hoc queries within the configured accuracy guarantees, with small memory requirements.
\end{abstract} 

\maketitle

\pagestyle{\vldbpagestyle}
\begingroup
\renewcommand\thefootnote{}\footnote{\noindent
This work is submitted for publication.\\
}\addtocounter{footnote}{-1}\endgroup


\input{1.intro}

\input{2.prelim}
\input{3.omni.s0}

\input{4.omni.s1}
\input{5.experiments}

\input{6.conclusions}


\bibliographystyle{ACM-Reference-Format}
\bibliography{references}

\end{document}

%% file: 1.intro.tex
\section{Introduction}
Filters and aggregates constitute the workhorse of data analytics, and are implemented in all databases. 
Accordingly, indexing and storage techniques have been implemented to answer such queries efficiently, even over big data. When it comes to streaming data that cannot be stored or real-time queried in its entirety, the go-to techniques are based on {\em sketches}~\cite{fntdbs12}: small-memory data structures that summarize the stream and can subsequently be used to execute aggregate queries. Example sketches from the literature support estimation of counts,  norms, and join aggregates~\cite{cormode2005,DBLP:journals/vldb/PapapetrouGD15,DBLP:journals/jcss/AlonMS99}), estimation of set sizes~\cite{fmsketch}, and identification of  frequent items and heavy hitters~\cite{DBLP:conf/icalp/CharikarCF02}. 

\begin{figure*}[ht]
\begin{minipage}[b]{0.49\linewidth}
\centering
    \includegraphics[width=\textwidth, trim = {8.5cm 15cm 9cm 1.5cm}]{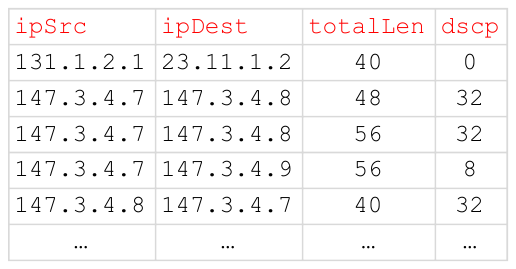}
    \cprotect\caption{A (simplified) stream, used as the running example. A sample query with multiple predicates on this stream may be, e.g., \verb|SELECT COUNT(*) FROM stream WHERE ipSrc=131.2.2.1 AND| \verb|ipDest=142.1.4.7 AND totalLen>40 AND dscp=0|}
    \label{fig:introfig}
\end{minipage}\hfill
\begin{minipage}[b]{0.49\linewidth}
    \centering
    \includegraphics[width=\textwidth, keepaspectratio]{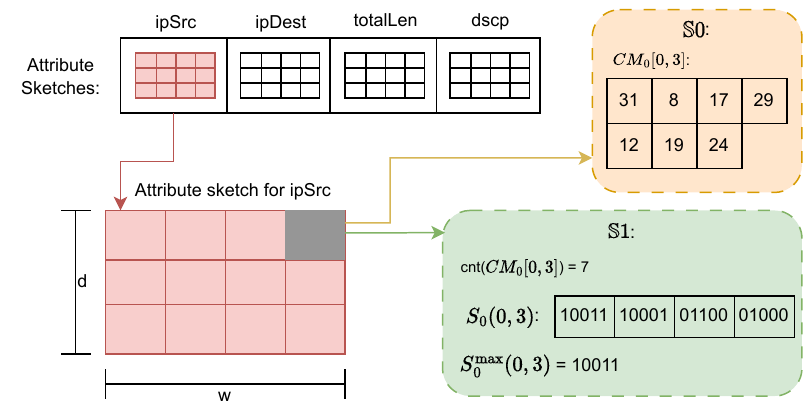}
    \caption{OmniSketch. The yellow-shaded box illustrates the contents of a cell in $\sketch0$. The green-shaded box corresponds to the 
    contents in $\sketch1$, for $B=4$ and signature size set to 5 bits.} 
    \label{fig:OmniSketch}
\end{minipage}
\end{figure*}

Despite being heavily used, most sketches to date focus on summarizing the frequency distribution based on a single attribute, or \emph{a pre-chosen combination} of attributes. Consider, for example, the domain of network monitoring, where the Count-min sketch finds frequent applications for statistics maintenance~\cite{cormode2005}. A standard IPv4 packet header defines at least 13 fields/attributes, including version,  header length, total length, the DSCP code point, source and destination address, protocol, and possibly one or more of 30 additional options. To be able to estimate the number of packets that satisfy a  combination of attribute values (defined at query time), we need to construct a sketch that uses as a key the combined values on these attributes (e.g., their concatenation). As a running example, consider the simplified IPv4 headers stream of Figure~\ref{fig:introfig}, which contains 4 of the 13 attributes. To summarize the distribution of the number of packets sent by each IP address, we need one sketch constructed using \verb+ipSrc+ as the key.  One additional sketch on \verb+ipDest+ is needed for summarizing the number of packets received by each IP address. If we also want to summarize the number of packets exchanged between any two IP addresses, we need to maintain a sketch that uses the concatenation of the source-destination IP addresses as the key.  The number of sketches that need to be maintained to support arbitrary predicates (in this example, any sub-combination of the 13 fields contained in the header) totals to $O(2^{13})$ -- the size of the powerset. Generalizing this to arbitrary use cases, estimating the frequency distribution for all sub-combinations of $p$ predicates requires maintenance of  $2^p-1$ sketches. This is clearly infeasible, both because of space requirements and strict efficiency constraints that arise in the context of data streams.~\footnote{
The recently proposed Hydra sketch~\cite{hydra} for multi-dimensional streams, addresses space constraints by hashing/summarizing the contents of all $2^p$ sub-sketches in the same sketch space. Still, as we also discuss later, Hydra's sketch maintenance complexity remains $O(2^p)$, which is simply not viable in real-world streaming use cases, where the number of attributes $p$ can be high (e.g., tens or hundreds).}

\vspace{.5em}
\noindent{\bf Our Contributions.}\/
In this work, we propose, analyze, and evaluate a novel sketching tool,
termed OmniSketch, that effectively addresses both space and time efficiency by combining sketching with sampling. OmniSketch, combines the compactness of sketches, which is necessary for reducing the memory constraints, with the generality of sampling, which is key for supporting general queries, on predicates that are dynamically decided at query time.  In a nutshell, an OmniSketch for summarizing a $p$-attribute data stream consists of $p$ individual small-memory sub-sketches, each similar to a Count-Min sketch. However, unlike Count-Min sketches, the cells in the OmniSketch sub-sketches contain fixed-size summaries of all records that hash into them. At query time, the sub-sketches that are relevant to the query, and the relevant cells from each sub-sketch, are located and queried to estimate the answers.  Unlike previous work, OmniSketch offers computational complexity (for both updates and queries) that scales linearly with the number of attributes -- instead of exponentially -- rendering it the only viable, general-purpose solution, to date, for summarizing fast-paced streams with many attributes in small space. Our sketch is backed by a theoretical analysis for providing formal error guarantees, and an automated initialization algorithm that builds on the theoretical analysis to fully utilize the available sketching memory.

We evaluate OmniSketch experimentally on both real and synthe\-ti\-cal\-ly-generated streams, and compare it with Hydra, the state-of-the-art competitor. Our experiments confirm that OmniSketch is the only viable option for summarizing complex streams, and comes with a favorable complexity-accuracy tradeoff.  In contrast, Hydra becomes extremely slow when summarizing streams with five or more attributes, and therefore fails to effectively summarize fast-paced streams.

\vspace{.5em}
\noindent{\bf Roadmap.}\/
The remainder of the paper is structured as follows. In Section~\ref{sec:prelim} we present the preliminaries and discuss the related work. In Section~\ref{sec:OmniSketch} we present OmniSketch and analyse its theoretical properties, whereas Section~\ref{sec:experiments}  summarizes our experimental results. We summarize the work and conclude with future plans in Section~\ref{sec:conclusions}. 

%% file: 2.prelim.tex
\section{Preliminaries and Related Work}
\label{sec:prelim}

\vspace*{.5em}
\noindent{\bf Preliminaries.}\/
OmniSketch inherits the basic structure of Count-Min
sketches~\cite{cormode2005}, and builds on the theory of k-minwise
hashing~\cite{pagh2014}. We will now briefly present these two works,
to the depth required for understanding our work.

\paragraph{Count-Min sketch.}
The Count-Min sketch, proposed by Cormode and Muthukrishnan in
2005~\cite{cormode2005}, became the de facto sketch for the summarization
of distributions of data streams.  A Count-Min sketch $CM$ is a
2-dimensional array of width $w$ and depth $d$, accompanied by $d$
pairwise-independent hash functions that map the input to the range
$\{1,\ldots w\}$.  Let $CM[j, h_j(\cdot)]$ denote the counter at
row $j$, column $h_j(\cdot)$ in the 2-dimensional array. 
A record $r$ (e.g., an IP address) is
added once to the sketch by increasing the count at $CM[j, h_j(r)]$ for
$j \in \{1 \ldots d\}$. The number of arrivals of any query $q$ in the  
stream  is estimated by finding the
corresponding cells $CM[j, h_j(q)]$ per row $j$ and returning the
minimum count, i.e.,
$\hat{f}(q) = \min_{j \in \{1 \ldots d\}}(CM[j, h_j(q)])$. 
Due to hash
collisions (items other than $q$ that fall in the same
counters) $\hat{f}(q)$ is potentially an overestimate of the true
frequency $f(q)$. By setting $w = \lceil e/\epsilon\rceil$ and
$d = \lceil \ln(1/\delta)\rceil$, we have that $\hat{f}(q) - f(q) \leq \epsilon N$ with
probability $\geq 1-\delta$, where $N$ is
the length of the stream.

Count-Min sketches also support range queries, by decomposing ranges to canonical
covers~\cite{cormode2005}.  Adding support for range queries for an
attribute $a$ increases the space and time complexity roughly by a factor of
$O(\log(|\mathcal{D}(a)|))$, where $\mathcal{D}(a)$ denotes the
attribute's domain.

Due to the attractive cost-accuracy tradeoff of Count-Min sketches, it is also 
possible to maintain multiple sketches on fast-paced data streams, each summarizing 
one attribute. For example, the stream of our running example 
(Figure~\ref{fig:introfig}) could be summarized on 4 individual Count-min sketches, 
thereby enabling frequency estimations with any one of the following four filtering conditions: 
\verb+WHERE ipSrc=?+, \verb+WHERE ipDest=?+,
\verb+WHERE totalLen=?+, \verb+WHERE dscp=?+ (with \verb+?+ denoting the predicate value).
However, Count-min sketches do not support multiple predicates in the same query, e.g.,  
\verb+WHERE ipSrc=?+ \newline
\verb+AND ipDest=?+. If these predicate combinations are known 
before observing the stream, then a single sketch can be built for each of these combinations,
indexing the concatenation of the attributes. For example, to support the query 
\verb+WHERE ipSrc=? AND +  \verb+ipDest=?+, we can construct a \emph{composite} key for each 
record \verb+<ipSrc,ipDest>+, and summarize these in a sketch. Then, 
queries that have both attributes in the predicates set can be answered by constructing 
the query's composite key in a similar 
way, and querying for it. However, if these predicate combinations are not known beforehand, or if 
we want to enable the user to query with arbitrary attribute combinations, 
summarization of a stream with $n$ attributes requires the 
construction of $2^n-1$ sketches, for covering all possible combinations of predicates.  
Therefore, this approach is not a viable solution, in terms of both time and space complexity.~\footnote{
Note that building a Count-Min sketch on the composite key comprising all $n$ attributes is also not a viable solution. 
Besides the obvious dimensionality curse problem for large $n$, such a sketching structure offers no space- or time-efficient way of handling 
``don't care'' attributes (for which no value constraints are specified) when querying with dynamic predicates on arbitrary attribute subsets. 
Possible approaches would still require a space/time blowup that is exponential in $n$.
}

\eat{
\todo{@Mino: the following is not needed/not helpful. Working out the details of this solution is a short paper itself, 
and it is a crazy solution. Anybody that could think of it would already dismiss it as non-viable.}
An  approach that avoids constructing one sketch for each possible combination 
is to construct a single Count-min sketch containing all attributes, and to 
enable range queries on all attributes of the sketch. Queries that contain all sketch attributes 
as predicates can be easily answered by concatenating all query's values to create a composite key. 
For queries that contain predicates only on some of the attributes, we can exploit the range query capabilities 
of the sketch to cover the whole domain of the attributes that are not bound to a value. This approach, however, comes with two key downsides, both stemming from the fact that each record is inserted 
in the sketch a logarithmic number of times \emph{per attribute}. 
First, it increases space and time complexity by a logarithmic factor for each stream's attribute. 
For example, the cost of summarizing four 32-bit attributes will increase by a factor of $\left(\log_2(32)\right)^4$.
Second, it increases the effective $\epsilon$ value of the sketch. Therefore, satisfying users' desired accuracy requirements requires appropriate scaling of the sketch's $\epsilon$ value, which leads to a further increase of the sketch size. 
}

\paragraph{K-minwise hashing.}
In this work, we rely on minwise hashing to estimate the cardinality of the intersection of 
multiple sets. The key idea behind all minwise hashing schemes is to hash all items 
in each set using one or more hash functions, 
and to keep only the $B$ items with the smallest hash values per hash function as samples of each set. 
The size of the intersection of these samples for the different sets can then be scaled to estimate the 
cardinality of the intersection of the sets. 

The K-minwise hashing algorithm that we will be using in this work estimates the sets intersection cardinality as follows~\cite{pagh2014}: Let $R_1, R_2, \ldots, R_p$ denote the $p$ sets for which we want to estimate the cardinality of their intersection. 
First, we construct the summary $S_i$ for each set $R_i$ by hashing each item with a global hash function  $g(\cdot) \rightarrow \{0,1\}^b$,  and retaining the $B$ smallest hash values, with $b$ set to at least $\lceil\log(4B^{2.5}/\delta)\rceil$. The cardinality of the intersection of $p$ sets $|R_\cap| = |R_1 \cap \ldots R_p|$ can then be 
estimated from these summaries as $\hat{R}_{\cap} = |\bigcap\limits_{i=1}^p S_i| *  n_{\max}/B$ with $ n_{\max} = \max_{i=1}^p|R_i|$.
This estimate comes with $(\epsilon, \delta)$ guarantees when $\hat{R}_\cap \geq 3 n_{\max}\log(2p/\delta)/(B\epsilon^2)$. 
When the above condition fails, a weaker bound can be shown: $0 \leq |R_\cap| \leq 4 n_{\max}\log(2p/\delta)/(B\epsilon^2)$, 
with probability $1-\delta\sqrt{B}$. 

We chose K-minwise hashing over other sampling methods, since this algorithm has been shown to perform equally well 
or outperform other sampling methods (including \cite{bloom, bbitminwise, onePerm}), and has space complexity that nearly 
matches the theoretical lower bound for the problem~\cite{pagh2014}. Furthermore, the chosen algorithm also works with streaming data, as the samples can be maintained incrementally.

\eat{
Note that K-minwise hashing, as well as other similar techniques for estimating set and set-intersection cardinality of sets and sets intersections that will be discussed in Section~\ref{sec:related} do not support predicates. Therefore, they cannot answer the queries that are the focus of this work (aggregates with predicates) out-of-the-box.
}

\paragraph{The stream model and supported queries.} 
The  input data is a stream of records, e.g., similar to the records of our running example of Figure~\ref{fig:introfig}. Let $\mathcal{A}=\{a_1, a_2, \ldots, a_{|\mathcal{A}|}\}$ denote the stream attributes (in the running example, ipSrc, ipDest, totalLen, dscp). Each record also comes with a unique record id (rid), denoted by $a_0$ -- if no record keys exist, such unique rids can be easily constructed at ingestion time (e.g., using an arrival counter). We assume a landmark stream query model~\cite{muthu:fnttcs05}, where a query is posed any time while the stream is ingested and  refers to all stream arrivals until that time.~\footnote{
We briefly discuss extending OmniSketch to handle general updates (i.e., record deletes as well as inserts) later in the paper.
}
Formally, let $\mathcal{R}=\{rec_1, rec_2, \ldots rec_N\}$ denote the stream arrivals up until query time. The query $q$ is a count query, comprising a conjunction of selection  predicates for \emph{any} subset 
$\{a_i, a_j, a_k, \ldots\}$ of $\mathcal{A}$:
$$
\text{Count}\left(rec \in \mathcal{R}~ |~ rec ~\text{satisfies}~ q = q_i \wedge q_j \wedge q_k \ldots \right)
$$
where each predicate $q_i$ can be a disjunction of range and equality predicates on an individual attribute $a_i$.

\vspace{.7em}
\noindent{\bf Other Related Work.}\/
The state-of-the-art sub-population sketch for multi-dimensional data streams is Hydra~\cite{hydra}.
At the outer layer, Hydra consists of a Count-Min sketch of width $w$ and depth $d$, accompanied with $d$ pairwise-independent hash functions that map the input domain to $[1\ldots w]$. Each cell in this sketch contains a nested universal sketch~\cite{universalSketch}, for keeping detailed statistics. When a record of $p$ attributes is read, it is hashed in the Hydra sketch as follows. First, all $2^p-1$ possible combinations of the record's attributes are computed. Each of these combinations defines a sub-population where this record belongs. Each combination is then hashed with the $d$ hash functions to find the corresponding cells in the outer sketch (one cell per row), and the combination is finally added in the contained universal sketches. For query execution, the query predicates are combined to create a key, which is then hashed using the same $d$ hash functions for finding the corresponding cells in the outer sketch. We then query the nested universal sketch using the same key, to estimate the frequency.

In a thorough experimental evaluation, Hydra is shown to outperform other approaches, and provides interactive query latency. However, the sketch has two critical downsides that make it a non-viable option for streams with many attributes. First, recall that each record is added $2^p-1$ times in the sketch, with different keys -- once per possible combination. As we will demonstrate later, this exponential complexity becomes a challenge already for a modest number of indexed attributes, i.e., $p=4$, in terms of time complexity, causing problems when summarizing fast-paced streams. Second, because of this sharp increase (exponential with $p$) of the number of additions to the sketch, the approximation error also rapidly increases. Our experimental results show that while Hydra demonstrates excellent performance for small $p$ values, its error (as well as update time) increase exponentially with $p$.

\paragraph{Techniques for set cardinality estimation.}
Distinct sampling was proposed for estimating the number of distinct items in a stream, based on a small-memory sample of the stream~\cite{distinctsampling}. The key idea is to maintain samples at different levels, with a total memory budget $B$. A record is included in each sampling level $l$ with a probability $2^{-l}$. When the total number of samples (across all levels) exceeds $B$, the lowest level is dropped, thereby releasing approximately half the memory and making space for more samples at higher levels. An $(\epsilon, \delta)$ estimate of the number of distinct items in the stream can be computed by multiplying the number of items in the smallest surviving sampling level $l_{min}$ with a scale factor $2^{l_{min}}$. Distinct sampling can also be used for estimating the size of set intersection, by exploiting coordinated sampling. In our context, distinct sampling could be used as an alternative of K-minwise hashing to progressively maintain stream samples with a fixed memory budget. In practice, K-minwise hashing was shown to better exploit the available memory in our experiments. Therefore, we do not report experiments with distinct sampling.

The 2-Level hash sketch (2LHS)~\cite{ggr:vldbj04} was proposed for cardinality estimation of arbitrary set expressions, on streams that contain general updates (record insertions and deletions). A 2LHS $\mathcal{X}_A$ comprises two levels of buckets and is implemented as a three-dimensional array of size $k \log(M) \times s \times 2$, where $k$ and $s$ are user-tunable parameters that control the estimation accuracy, and $M$ is the domain size of the input.
Conceptually, the 2LHS can be viewed as a generalization of distinct sampling that can be employed to give $(\epsilon,\delta)$ cardinality estimates of set unions, intersections, and differences over general update (i.e., turnstile~\cite{muthu:fnttcs05}) data streams.
While 2LHS could be used as an alternative for K-minwise hashing, there is added space/time complexity in the 2LHS structure, which is actually necessary for handling record deletions in the stream.
Thus, we focus our OmniSketch design on K-minwise hashing which offers a much simpler and more space-efficient solution for insert-only streams, and discuss possible extensions for general updates later in the paper.

\eat{
The first level of buckets has a total size of $k \log(M)$, and it is accompanied by an LSB-based hash function $h(\cdot)$ that maps each input item $x$ to a value $i \in [0,k \log(M)-1]$ with probability $2^{-i}$, similar to the hashing for FM-sketches~\cite{fmsketch}. The second level consists of an array of size $s\times 2$, and a family $\mathcal{G}$ of $s$ hashes $g_1(\cdot), g_2(\cdot),\ldots g_s(\cdot)$ that uniformly map the input to 0 or 1. Any new arrival $x$ is first hashed with the first-level hash function $h$ to find the corresponding first-level bucket. Then, the input is hashed with all hashes $g_i \in \mathcal{G}$, and the counter at the respective cell $\mathcal{X}_A[h(x), i, g_i(x)]$ is increased by one. The structure can be used for estimating cardinalities of set unions, intersections, and differences, with $(\epsilon,\delta)$ guarantees. Similar to distinct sampling, the 2LHS could also be used as an alternative to K-minwise hashing in OmniSketch. However, 2LHS focuses on multisets, and therefore stores additional information. In our case, each item needs to be added to the set exactly once.  Therefore, K-minwise hashing offers better space utilization.
}

A number of other sampling-based techniques have been proposed in the literature for estimating the cardinalities of set unions, set intersections, and arbitrary set expressions over (insert-only) record streams~\cite{DBLP:conf/icdt/0001LRT16, DBLP:journals/pvldb/Ertl21, bbitminwiseHashing, Lemiesz-OnAlgebraDataSketches, lemiesz-efficientFramework}. At their core, these methods are similar to K-minwise hashing, with similar complexities and theoretical guarantees. In principle, any of these could also be used with OmniSketch to construct the nested sketch. It is important to note, however, that none of these methods (including K-minwise hashing) provides support for predicate-based filtering. Consequently, they cannot be directly utilized to estimate  answers to aggregate queries with predicates, which is the main focus of our work.

%% file: 3.omni.s0.tex
\section{OmniSketch: estimating frequencies with arbitrary predicates}
\label{sec:OmniSketch}

We now present OmniSketch, a sketch that allows efficient estimation of queries with predicates over fast-paced streams. We  gradually construct the final sketch, in three successive steps. Initially, we describe an extension of the standard Count-Min sketch, termed $\sketch0$, that maintains additional data in the cells. This additional information is leveraged for estimating cardinalities involving predicates using a straightforward generalization of the Count-Min estimator. 
Then, keeping the $\sketch0$ sketch structure unchanged, we present an improved estimation algorithm and the corresponding theoretical analysis to tighten the error bounds, at no extra complexity. We refer to the new sketch estimator as $\sketch0_{\cap}$, to distinguish from the earlier estimator (termed $\sketch0_{\min}$). Finally, we present the ultimate OminiSketch ($\sketch1$) which merges $\sketch0_{\cap}$ with a sampling technique to guarantee sublinear space complexity. Unlike previous works, all three sketches allow very efficient updates (with logarithmic complexity), and can therefore easily handle fast-paced streams.

All three sketches are based on a common architecture: each comprises a collection of sub-sketches, with one sketch assigned to each searchable attribute, i.e., an attribute for which predicate support is desired. We refer to the sub-sketches as attribute sketches, and denote them by $CM_1, CM_2, \ldots CM_{|\mathcal{A}|}$, where $\mathcal{A}$ corresponds to the set of searchable attributes. Each $CM_i$ is an array of size $w \times d$, accompanied with $d$ pairwise-independent hash functions ${h_i^1(\cdot),h_i^2(\cdot),\ldots h_i^d(\cdot)}$ (similar to Count-Min sketches). However, unlike traditional CM-sketches, each cell in $CM_i$ contains a list of record ids (rids), or their fingerprints. Rids are placed/hashed in each sketch based on the record value on the corresponding attribute. During query execution, the attribute sketches corresponding to the query's predicates are queried, and the record ids retrieved by these are intersected, to compute an estimate of the answer. Figure~\ref{fig:OmniSketch} illustrates the general sketch architecture, for a sample dataset with searchable attributes: <ipSrc, ipDest, totalLen, dscp>. 
The difference between the three sketches relates to: (a) the way the record ids are sampled and stored in the attributes sketches, and (b) the theory backing the estimators which affects the tightness of the bounds. 

In the ensuing discussion, we use $\mathcal{A}=\{a_1, a_2, \ldots, a_{|\mathcal{A}|}\}$ to denote the set of searchable attributes, $q$ to denote a query that contains $p$ predicates, and $q_i$ to denote the predicate for attribute $a_i \in \mathcal{A}$. For ease of presentation, unless otherwise mentioned, we assume that each predicate $q_i$ is an equality predicate on attribute $a_i$ with a constant value $v_{q_i}$; that is, $q_i:= (a_i = v_{q_i})$.
The domain of each attribute $a_i$ is denoted by $\mathcal{D}(a_i)$. 

Without loss of generality, we assume that $q$ contains all attributes of $\mathcal{A}$, i.e., $p=|\mathcal{A}|$ (if some of the attributes are not contained in the query, we simply ignore the corresponding sketches during query execution); in general, $p\leq|\mathcal{A}|$. We summarize frequently used notation in \autoref{tab:notation}.

\begin{table}[h]
    \centering
    \resizebox{\columnwidth}{!}{%
    \begin{tabular}{l|l}
\hline
Notation       & Description                                                    \\ \hline
N              & Stream length.                                                 \\
f(q)           & Number of records satisfying query q                           \\
d              & Sketch depth (number of rows)                                  \\
w              & Sketch width (number of columns)                                \\
B              & Maximum sample size per cell                                    \\
\hline
$CM_i$   & Sketch for attribute $i$.          \\
$CM_i[j, k]$   & Cell at row $j$, column $k$, of sketch $CM_i$.          \\
$h^j_i(\cdot)$ & Hash function for attribute sketch, row $j$ of attribute $i$.      \\
$R(CM_i[j, k])$    & Set of records hashed to cell $CM_i[j, k]$.                       \\
$S_i(j, k)$    & Set of records in the sample of cell $CM_i[j, k]$.                    \\
$C(q)$         & Set of cells $CM_i[j, h^j(v_{q_i})]$ accessed to answer the query.    \\
$R_{\cap}$     & Set of records hashed to all cells in $C(q)$.                     \\
$n_{\max}$     & Max. number of records hashed to any cell $CM_i[j, k] \in C(q)$.     \\ \hline
\end{tabular}
}
    \caption{Frequently used notation.}
    \label{tab:notation}
\end{table}

\subsection{Sketch $\sketch0$: Count-Min with rid-sets for estimating queries with predicates}
\label{sec:version0}

\paragraph{Initialization}\label{sec:construction}
At initialization time, we choose uniformly at random $|\mathcal{A}| \times d$ pairwise-independent hash functions $h^1_1(\cdot),$ $ h^1_2(\cdot),$ $ \ldots, h^d_{|\mathcal{A}|}(\cdot)$, with $h_i^j(): \mathcal{D}(a_i) \rightarrow \{1,\ldots, w\}$. We also construct $|\mathcal{A}|$ arrays $CM_1,$ $CM_2 \ldots, CM_{|\mathcal{A}|}$, of size $w \times d$, and initialize each of their cells to contain an empty linked list. In order to get $(\epsilon, \delta)$-guarantees on the estimate, we set $w = 1 + \lceil e/\epsilon \rceil$ $=\Theta(1/\epsilon)$ and $d = \lceil \ln(1/\delta)\rceil$.

\paragraph{Insertion}\label{sec:updating}
A new record $r$ needs to be added to all $|\mathcal{A}|$ sketches. We use $r_i$ to denote the value of record $r$ on attribute $a_i$, and $r_0$ to denote its unique record id (rid). For each searchable attribute $a_i$ and for each row $j=\{1,\ldots, d\}$, we add $r_0$ to all linked lists at positions $CM_i[j, h_i^j(r_i)]$.

Note that, as defined, the $\sketch0$ structure is not strictly a ``sketch'' since, by storing complete rid-sets, it requires space that is linear in the stream length $N$. The only summarization $\sketch0$ performs is by ``blurring'' individual attribute values through hashing into Count-Min buckets. 
Still, it provides a conceptually useful first step towards our final OmniSketch solution.

\vspace*{.5em}
\noindent {\bf The $\sketch0_{\min}$ Estimator.}\/
Let $f(q)$ denote the number of records satisfying all predicates in $q$. Following the conventional Count-Min estimation procedure~\cite{cormode2005}, we can compute an estimate $\hat{f}(q)$ as follows: for each row $j=\{1,\ldots, d\}$, we compute the size of the intersection of all records stored in cells $CM_1[j, h_1^j(v_{q_1})], CM_2[j, h_2^j(v_{q_2})],$ $ \ldots,$ $ CM_p[j, h_p^j(v_{q_p})]$. We repeat this for the $d$ rows, and return the minimum value as an estimate.
Formally: 
\begin{equation}\label{eq:estimate_ver0}
    \hat{f}(q) = \min_{j=\{1\ldots d\}} \left|\bigcap\limits_{q_i \in q} CM_i[j, h^j(v_{q_i})]\right|
\end{equation}

Intuitively, the intersection will contain all records $r$ that satisfy the query, since these will always hash at the same cell as the query predicate, for all predicates, However, the intersection may also contain some false positives, i.e., records that were hashed in $CM_i[j, h_i^j(v_{q_i})]$ due to one or more random collisions. By taking the minimum intersection size across all $d$ rows, we try to minimize the effect of  such false positives.

\paragraph{Error bounds.}
The following lemma provides probabilistic guarantees for the estimator of Eqn.~\ref{eq:estimate_ver0}.
\begin{lemma}\label{lemma:simple_ver0}
    Let $\hat{f}(q)$ be the estimate provided by \autoref{eq:estimate_ver0} on sketches constructed with $d = \lceil\ln(\frac{1}{\delta})\rceil$, $w = 1 + \lceil\frac{e}{\epsilon}\rceil$. 
    For query $q$ with $p$ predicates:
    \begin{equation}\label{eq:lemma_wo_ass}
        Pr\left(|\hat{f}(q) - f(q)| \leq \epsilon N \right) \geq 1-\delta
    \end{equation}
    
\end{lemma}

\begin{proof}

Consider row $j \in \{1\ldots d\}$. Let us define an indicator variable $I_{y,j}$, which takes the value of 1 for all records $y$ satisfying the following condition: 

\[
\forall_{q_i\in q}[h_i^j(v_{q_i}) = h_i^j\big(y_i\big)] \wedge \exists_{q_i \in q}[v_{q_i} \neq y_i] 
\]

Informally, the described condition is true for all false positive records, i.e., records that are contained in the intersection but do not fully satisfy the query.
Also, variable $X^j=\sum_{y=1}^{N - f(q)} I_{y,j}$ is a counter of these false positives for row $j$.

Recall that the estimator for row $j$ is as follows: $\hat{f}^j(q) = |R_\cap^j|$, where $R_\cap^j$ denotes the intersection of the cells at position $h_i^j(v_{q_i})$, for $i=\{1,\ldots, p\}$. 
We need to prove that: (a) all records that satisfy all predicates will be contained in $R_\cap^j$, i.e., there will be no false negatives, and (b) the total number of false positives $X$ is upper-bounded w.h.p..

For (a), notice that for any record $y$ satisfying $y_i=v_{q_i}$ for $i=\{1\ldots p\}$, we will have $y_i=v_{q_i} \Rightarrow h_i^j(y_i)=h_i^j(v_{q_i})$. Therefore, all records satisfying all predicates will be included in $R_\cap^j$.

For (b), a record will be a false positive and increase the value of $X^j$ by 1 if for all attributes $y_i$ for which $y_i \neq v_{q_i}$, we have $h_i^j(y_i)=h_i^j(v_{q_i})$. Let $X_k^j$ denote the number of records contained in $R_\cap^j$ that satisfy \emph{exactly} $p - k$ predicates, i.e., they do not satisfy $k$ of the $p$ predicates, but they are retrieved by the algorithm due to hash collisions. Then, $X^j=\sum_{k=1}^p {X^j_k}$. 

By construction of the hash functions, a record $y$ with $y_i \neq v_{q_i}$ will have $h_i^j(y_i)=h_i^j(v_{q_i})$ with probability $1/w$. Since the hash functions are pairwise-independent, the probability that a record differing from $q$ in $k$ attributes hashes in the same cells is $1/w^k$.
Therefore, $E[X^j_k] \leq (N - f(q))/w^k$, and 
$E[X^j]=\sum_{k=1}^p E[X^j_k] \leq (N - f(q)) * (1/w + 1/w^2 + \ldots + 1/w^p) = (N - f(q)) * (1 - (1/w)^p) / (w - 1)$. By setting $w = 1+\lceil e/\epsilon\rceil$ we get $E[X] = (N - f(q))/e*(1-\epsilon^p)/\epsilon \leq \epsilon (N - f(q))/e \leq \epsilon N/e$.

Notice that we have $d$ rows, each with different hash functions. By Markov's inequality, we have 
\begin{align*}
    Pr[X \geq \epsilon (N - f(q))] &= Pr[\forall_{j \in [1\ldots d]}.~~ \hat{f}^j(q) > f(q) + \epsilon N] \\
    &=Pr[\forall_{j \in [1\ldots d]}.~~ X^j > e E[X]] \\
    &< e^{-d}
\end{align*} 
where $f(q)$ denotes the true answer and $X^j$ denotes the false positives at row $r$, i.e., $X_j=\hat{f}^j(q)-f(q)$. The final bound
follows by setting $d=\lceil \ln(1/\delta)\rceil$.
\end{proof}

\vspace{.5em}
\noindent {\bf $\sketch0_{\cap}$:~~An Improved Estimator.}\/
The $\sketch0_{\min}$ estimator is based on the conventional estimation logic in the standard Count-Min sketch~\cite{cormode2005}; that is, it produces an estimate per row and takes the minimum across all the $d$ row estimates.
Similar to standard Count-Min estimation, it is easy to see that $\sketch0_{\min}$ can only overestimate the true count due to hash bucket collisions (false positives). 
However, compared to the standard Count-Min, each bucket in $\sketch0$ contains much more detailed information that can be exploited to provide much tighter estimates (i.e., with less false positives).
The key observation here is that, by construction of $\sketch0$, 
each record that satisfies the full query will end up in the cells of \emph{all} $d$ rows for this query and for all $p$ predicates, whereas false positives are expected to only end up in a few of the rows. 
Thus, without modifying the sketch construction or space/time complexity, we can get a tighter estimator by removing the $\min$ operation across rows and simply intersecting the rid-sets  returned by all $d$ rows. We refer to this tighter $\sketch0$ estimator as $\sketch0_{\cap}$, and formally define it as follows:
\begin{equation}\label{eq:estimate_ver1}
    \hat{f}(q) = \left| \bigcap\limits_{i = \{1\ldots p\}, j = \{1\ldots d\}} CM_{i}[j, h_i^j(v_{q_i})]\right|
\end{equation}

Clearly, this new $\sketch0_{\cap}$ estimator can also only overestimate the true count due to hash collisions -- at the same time, it guarantees fewer false positives since it is always less than or equal to the $\sketch0_{\min}$ estimate in Eqn.~\ref{eq:estimate_ver0}. 
We now demonstrate the stronger error guarantees of $\sketch0_{\cap}$, proving 
that it allows us to bound the error by $\epsilon^d (N - f(q))$ instead of $\epsilon (N - f(q))$, without changing the space complexity of the sketch.
\begin{lemma}\label{lemma:simple_ver1}
    Let $\hat{f}(q)$ be the estimate provided by \autoref{eq:estimate_ver1} on sketches constructed with $d = \lceil\ln(\frac{1}{\delta})\rceil$, $w = 1 + \lceil\frac{e}{\epsilon}\rceil$. For query $q$ with $p$ predicates:
    \[
    Pr\left(|\hat{f}(q) - f(q)| \leq \epsilon^d (N-f(q)) \leq \epsilon^d N \right) \geq 1-\delta
    \]
\end{lemma}

\begin{proof}
The proof is similar to the proof for Lemma~\ref{lemma:simple_ver0}.

We define an indicator variable $I_y$, which takes the value of 1 for all records $y$ satisfying the following condition:
\[\forall_{q_i\in q}[\forall_{1 \leq j \leq d} [h_i^j(v_{q_i}) = h_i^j\big(y_i\big)] \wedge \exists_{q_i \in q}[v_{q_i} \neq y_i]]\]

Informally, $I_y$ becomes 1 for all false positive records, i.e., the records that hash to the same cells as the query predicate value, at \emph{all} $d$ rows and at \emph{all} $p$ predicates. 
Also, $X=\sum_{y=1}^{N - f(q)} I_y$ is a counter of the total number of false positive records.

We will prove that (a) all records that satisfy all predicates will be contained in $\bigcap\limits_{j=1}^d R_\cap^j$, i.e., there will be no false negatives, and, (b) the number of false positives $X$ is upper-bounded w.h.p..

For (a), notice that for any record $y$ satisfying $y_i=v_{q_i}$ for $i=\{1\ldots p\}$, we will have $y_i=v_{q_i} \Rightarrow \forall_{1\leq j \leq d}[h_i^j(y_i)=h_i^j(v_{q_i})]$. Therefore, all records satisfying all predicates will be included in the intersection of all $R_\cap^j$, denoted as $R_\cap$.

For (b), a record $y$ will increase the value of $X$ if for all attributes $y_i$ for which $y_i \neq v_{q_i}$, and for all rows $j \in \{1\ldots d\}$, it collides with $h_i^j(v_{q_i})$, i.e., $h_i^j(y_i)=h_i^j(v_{q_i})$.
The probability of this collision at a single row $j$ is $Pr[h_i^j(y_i)=h_i^j(v_{q_i})]=1/w$. Furthermore, by independence of the hash functions across sketch rows, the probability that a record collides with the query at all $d$ rows $j$ will be $1/w^d$.

Let $X_k$ denote the number of records from $R_\cap$ that satisfy \emph{exactly} $p - k$ predicates. Then, $E[X_k] \leq (N - f(q))/w^{kd}$, and $E[X] = \sum_{k=1}^p E[X_k] \leq (N - f(q)) \sum_{k=1}^p 1/w^{kd} = (N - f(q)) \frac{1 - w^{-d(p+1)}} {w^d  - 1} < (N - f(q))/(w^d - 1)$. By setting $w= 1+\lceil e/\epsilon \rceil$, we get $E[X] \leq (N - f(q)) / ((1+e/\epsilon)^d - 1) \leq \epsilon^d (N - f(q)) / e^d$.

The following bound follows directly from Markov inequality:
\begin{align*}  
    Pr[X \geq \epsilon^d (N - f(q))] &= Pr[X \geq e^d E[X]] \\
    & \leq e^{-d}
\end{align*}
Then, by setting 
$d=\lceil \ln(1/\delta)\rceil$ we get our final bound, 
$Pr[X \geq \epsilon^d (N - f(q))] \leq \delta \Rightarrow 
Pr[X \geq \epsilon^d N] \leq \delta$
\end{proof}

%% file: 4.omni.s1.tex
\subsection{Sketch $\sketch1$ (OmniSketch): Achieving sublinear space through sampling}\label{sec:sketch1}
Sketch $\sketch0$  stores the record ids of all records, resulting in a space complexity of $O(d\times |\mathcal{A}|\times N)$, which is not viable for large data streams. Our full-fledged OmniSketch (also denoted by $\sketch1$ in what follows), achieves space complexity strictly sublinear in $N$ by taking and maintaining \emph{samples} of rids in the cells of the attribute sketches, using a K-minwise hashing algorithm~\cite{pagh2014}. We now  describe our OmniSketch solution, assuming that two key parameters (the maximum sample size per cell $B$ and the range of the sampling hash function $[0, 2^b-1]$) are already set. We explain how these values are determined later in this section. 

Similar to $\sketch0$, $\sketch1$ is composed of a collection of $|\mathcal{A}|$ attributes sketches. Furthermore, $\sketch1$ incorporates a hash function 
$g: \mathcal{D}(r_0) \rightarrow \{0, 1\}^b$, which maps each record id to a bit string of length $b$, with $b$ set to at least $\lceil\log(4B^{2.5}/\delta)\rceil$. 
Function $g$ is necessary for K-minwise hashing (Section~\ref{sec:prelim}). The $|\mathcal{A}|$  
attributes sketches $CM_1, CM_2, \ldots CM_{|\mathcal{A}|}$ are all of the same size $w \times d$. 
Each sketch cell $CM_i[j, k]$ contains: (a) the count of all items that were hashed in this cell, denoted as $cnt(CM_i[j, k])$, (b) a minwise 
sample $S_i(j, k)$ of maximum size $B$, of the rid hash values $g(r_0)$ of all records $r$ that were hashed in this cell, and, 
(c) the maximum hash value of all items contained in $S_i(j, k)$, denoted as \simax. 
Notice that $cnt(CM_i[j, k])$ also includes items that were hashed in $CM_i[j, k]$ but 
did not end up in the sample, and it is expected to be much greater than $B$ in a populated sketch. 
Fig.~\ref{fig:OmniSketch} (the green-shaded box) shows an example of a populated cell's contents, with $B=4$ and $b=5$.

\paragraph{Initialization} The user chooses the desired error guarantees $(\epsilon,\delta)$, with $\epsilon<0.25$. 
Let $(\epsilon_1, \delta_1)$ and  $(\epsilon_2, \delta_2)$ correspond to the $(\epsilon, \delta)$ configurations of the minwise sampling algorithm and the attribute sketch respectively, and $\epsilon_1=\epsilon$, $\epsilon_2=(\epsilon/(1 + \epsilon))^{1/d}$, and $\delta_1=\delta_2=\delta/2$. We initialize all attribute sketches, by choosing $w$ and $d$ similar to $\sketch0$.  Each cell in these 
sketches is initialized with an empty sample $S_i(j, k)$, and with $cnt(CM_i[j, k])=0$ and $\simax=\infty$.

\paragraph{Insertion}
To add a new record $r$ to the sketch, we first locate all cells across the $|\mathcal{A}|$ sketches that correspond to the values of $r$ (lines 2-4, Alg.~\ref{alg:updateOmni}). 
These are the cells $C(r) = \{CM_i[j, h_i^j(r_i)] : i=\{1,\ldots, |\mathcal{A}|\}, j=\{1,\ldots, d\}\}$. For each of these cells $CM_i[j,k] \in C(r)$, 
we increase $cnt(CM_i[j, k])$ by one (line 5). Then, we examine whether $g(r_0)$ should be added to the sample of the cell (lines 6-10), as follows.
If the sample contains less than $B$ items, then $g(r_0)$ is added to it.
If, on the other hand, $|S_i(j, k)| = B$ and $g(r_0) < \simax$ (i.e., the sample is full, but the new record's id has a smaller hash value $g$ compared to 
another record from the sample),  we remove  $\simax$ from the sample to make space for $g(r_0)$. Finally, we add  $g(r_0)$ to $S_i(j, k)$, and 
recompute \simax.~\footnote{To speed-up the computation of \simax, as well as the estimation process, the samples $S_i(j, k)$ are maintained in a red-black tree.
Therefore, re-computation of \simax takes (amortized) constant time.}
The complexity of inserting an element is $O(|\mathcal{A}|\times d\times \log(B))$.

\begin{algorithm}
\caption{Adding a record $r$ to the sketch}\label{alg:updateOmni}
\begin{algorithmic}[1]
\item \textbf{Input:} record $r$
\For{$i \in \{1 \ldots |\mathcal{A}|\}$}
    \For{$j \in \{1 \ldots d\}$}
    \State $k=h_i^j(r_i)$
    \State $cnt(CM_i[j, k])$ ++
    \If{$|S_i(j, k)| < B$}
        \State Add $g(r_0)$ to $S_i(j, k)$
        \State Recompute \simax
    \EndIf
    \If{$|S_i(j, k)| == B$} 
        \If{$g(r_0) < $\simax}
            \State Remove \simax from $S_i(j, k)$
            \State Add $g(r_0)$ to $S_i(j, k)$
            \State Recompute \simax
        \EndIf
    \EndIf
    \EndFor
\EndFor
\end{algorithmic}
\end{algorithm}

\vspace{.5em}
\noindent {\bf OmniSketch Query Estimation.}\/
Let $q$ denote the input query. As with $\sketch0$, we assume that $q$ contains all $\mathcal{A}$ attributes, i.e., $p=|\mathcal{A}|$; if an attribute is not contained in $q$, the estimator simply ignores the corresponding attribute sketch. Each query attribute value $v_{q_i}$ is
hashed using the $d$ hash functions of the corresponding sketch $CM_i$, which
 leads to $d$ cells per sketch. Let $C(q)$ denote the set of cells 
across all sketches that are accessed to answer
the query, i.e., $C(q)=\{CM_i[j,h^j(v_{q_i})]: i=\{1\ldots p\}, j=\{1\ldots d\}\}$, and $n_{\max}=\max_{CM_i[j,k] \in C(q)} cnt(CM_i[j,k])$ be the maximum count of these cells.
Following the reasoning of our improved $\sketch0_{\cap}$ estimator, let 
\[S_\cap = \bigcap\limits_{i = \{1\ldots p\}, j = \{1\ldots d\}} S_i(j, h^j(v_{q_i}))
\]
denote the 
intersection of all samples stored in all cells in $C(q)$.
Our OmniSketch  estimator $\hat{f}(q)$ is computed as follows:

\begin{equation}\label{eq:estimate_ver2}
    \hat{f}(q) = \frac{n_{\max}}{B} \times \left| S_\cap \right|
\end{equation}

Intuitively, this estimator computes the cardinality of the intersection of all samples in $C(q)$ and scales it 
up by a factor of $\frac{n_{\max}}{B}$, to account for the sampling. 
Based on the analysis of K-minwise hashing~\cite{pagh2014}, our estimator 
comes with a sanity bound of $\frac{3n_{\max}\log(4pd\sqrt{B}/\delta)}{B\epsilon^{2}}$ to cover the 
case of insufficient samples in the intersection.

\paragraph{Efficient computation of the intersection.} Computing  $\left| S_\cap \right|$ requires a multi-way join over all samples that participate in the query. For large values 
of $B$, a naive hash-based computation of this intersection can take hundreds of milliseconds. To speed up query 
execution, we exploit the fact that the samples at each cell are already stored in a red-black tree, which 
allows for sorted iteration and search with complexity logarithmic in $B$. Our code is an extension of 
the sort-merge join for multi-way joins. Starting from the first cell, we get the first sample (the one with 
the minimum hash value), and execute a lookup on all other cells to check whether this hash value is contained 
in their samples. As soon as we find a cell that does not contain this sample, we get the smallest sample from that cell with 
a hash value greater than the one that failed, and resume our search from this value (no other value in between 
the failed and this value could be part of the join).
An interesting observation, 
which also becomes obvious in our experimental results (see Section~\ref{sec:experiments}), is that this algorithm 
typically becomes more efficient when answering queries with more predicates,  because it allows 
for larger steps. For example, if one of the cells contains very few samples, this cell will lead to skipping many candidate records at all other cells. This can bring the complexity of query execution from $O(p*d*B*\log(B))$ down to 
effectively $O(B\log(B))$.

\paragraph{Derivation of the error bounds.} 
$\hat{f}(q)$ has two sources of errors: (a) underestimation or overestimation due to K-minwise hash sampling, and, (b)
(one-sided) overestimation due to hash collisions in the outer Count-Min structure. We first provide a bound on the sampling error, which is oblivious to 
hash collisions, and then integrate the error due to hash collisions. Recall that  $(\epsilon_1, \delta_1)$ and 
$(\epsilon_2, \delta_2)$ correspond to the $(\epsilon, \delta)$ configurations of the K-minwise sampling algorithm
and the attribute sketch respectively.

We use $R(CM_i[j,k])$ to denote all records that are hashed into $CM_i[j,k]$, even if these do not end up in the sample, and $R_\cap = \bigcap\limits_{CM_i[j,k]\in C(q)} R(CM_i[j,k])$ to denote the intersection of records that are hashed in all cells in $C(q)$. Finally, $S_i(j, k)$ denotes the samples collected at cell $CM_i[j,k]$ and $S_\cap$ the intersection of samples of all cells in $C(q)$.

The following theorem and corollary provide the error guarantees for our OmniSketch estimator. (Note that in any realistic setting, $n_{\max} << N$.) 
\begin{theorem}
Consider an OmniSketch with $\epsilon_1=\epsilon$, $\epsilon_2=\big(\epsilon/(1 + \epsilon)\big)^{1/d}$, 
and $\delta_1=\delta_2=\delta/2$.  If 
$\left| S_\cap \right| < \frac{3\log(4pd\sqrt{B}/\delta)}{\epsilon^{2}}$, 
setting $\hat{f}(q) =$ $\frac{2n_{\max} \log(4pd\sqrt{B}/\delta)}{B \epsilon^2}$
satisfies $|f(q)-\hat{f}(q)| <$ $\frac{2n_{\max} \log(4pd\sqrt{B}/\delta)}{B \epsilon^2}$ with probability at least $1-\delta/2$. Otherwise, 
setting 
$\hat{f}(q)=$ $\frac{n_{\max}}{B} \times \left| S_\cap \right|$ satisfies
$|f(q)-\hat{f}(q)| \leq  \epsilon N$ with probability at least $1-\delta$.
\end{theorem}

\begin{proof}
We distinguish two cases, based on the value of $\left| S_\cap \right|$:

\textbf{Case 1} 
We first examine the case that $\left| S_\cap \right| \leq  \frac{3\log(2pd \sqrt{B}/\delta_1 )}{\epsilon_1^{2}}$.
From Theorem 4 of~\cite{pagh2014}, and by setting the theorem's $\delta$ to $\delta_1/\sqrt{B}$, we derive that  
$0 \leq |R_{\cap}| \leq \frac{4n_{\max} \log(2pd\sqrt{B}/\delta_1)}{B \epsilon_1^2}$. We also know that $f(q) \leq |R_\cap|$, since the latter may contain false positives -- records that do not fully satisfy all query predicates, but still end up in all cells of $C(q)$ due to hash collisions.  Therefore, $f(q) \leq \frac{4n_{\max} \log(2pd\sqrt{B}/     \delta_1     )}{B \epsilon_1^2}$. By returning 
$\hat{f}(q) = \frac{2n_{\max} \log(2pd\sqrt{B}/\delta_1)}{B \epsilon_1^2}$ we guarantee that $|\hat{f}(q)-f(q)| \leq \hat{f}(q)$ with probability $1-\delta_1$. Furthermore, setting $\delta_1=\delta/2$ and $\epsilon_1=\epsilon$ we get the final estimator for this case: $\hat{f}(q) = \frac{2n_{\max} \log(4pd\sqrt{B}/\delta)}{B \epsilon^2}$, and 
$|\hat{f}(q)-f(q)| \leq \frac{2n_{\max} \log(4pd\sqrt{B}/\delta)}{B \epsilon^2}$ with probability $\geq 1-\delta_1=1-\delta/2$.

\textbf{Case 2}
Now consider the case that $\left| S_\cap \right| >  \frac{3\log(2pd\sqrt{B}/\delta_1)}{\epsilon_1^{2}}$. 
From Theorem 4 of~\cite{pagh2014}:
\begin{equation} \label{eqn:boundSampling}
    \frac{n_{max}}{B} |S_\cap| = (1 \pm \epsilon_1) |R_\cap|
\end{equation}
with probability at least $1-\delta_1$. Notice however that due to hash collisions, not all records contained in $R_\cap$ (and hence, also in $S_\cap$) will fully satisfy the query. The number of false positives in $R_\cap$ can be estimated using Lemma~\ref{lemma:simple_ver1}: $|FP| = ||R_\cap| - f(q)| \leq \epsilon_2^d (N - f(q))$, with probability $1-\delta_2$. 

From Eqn.~\ref{eqn:boundSampling} and the triangle inequality we have 
\begin{align*}
\left|\frac{n_{max}}{B} |S_\cap| - f(q)\right| \leq & \left|\frac{n_{max}}{B} |S_\cap| - |R_\cap|\right| + (|R_\cap|-f(q))\\
\leq & \epsilon_1 |R_\cap| + \epsilon_2^d(N-f(q))\\
\leq & \epsilon_1 \bigg(f(q) + \epsilon_2^d(N-f(q))\bigg)    + \epsilon_2^d(N-f(q))
\end{align*}
with probability at least $1-\delta_1-\delta_2$.

Let us set $c=f(q)/N$ for convenience. Then: 
\begin{align*}
|\frac{n_{max}}{B} |S_\cap| - f(q)| \leq & \epsilon_1 \bigg(c N + \epsilon_2^d(N-cN)\bigg) + \epsilon_2^d(N-cN) \\
\leq & N [ \epsilon_1 c + \epsilon_1 \epsilon_2^d (1-c) + \epsilon_2^d(1-c)] \\
= & N \left[ \epsilon_2^d ( \epsilon_1 +1) - c( \epsilon_1\epsilon_2^d + \epsilon_2^d - \epsilon_1) \right]
\end{align*}

By setting $\epsilon_1=\epsilon < 1/4$, $\epsilon_2=\big(\epsilon/(1 + \epsilon)\big)^{1/d}$, $\delta_1=\delta_2=\delta/2$,  we get
$|\frac{n_{max}}{B} |S_\cap| - f(q)| \leq \epsilon N$, with probability $1-\delta_1-\delta_2$.
\end{proof}

The following corollary simplifies the estimation procedure, by always using the estimator proposed for case 2.
\begin{corollary}
Let $\hat{f}(q) = \frac{n_{\max}}{B} \times \left| S_\cap \right|$.~~ If $|S_{\cap}| < \frac{3n_{\max}\log(4pd\sqrt{B}/\delta)}{B\epsilon^2}$, then
$|f(q)-\hat{f}(q)| \leq \frac{4n_{\max}\log(4pd\sqrt{B}/\delta)}{B\epsilon^2}$ with probability at least $1-\delta/2$. Otherwise, the same estimator satisfies $|f(q)-\hat{f}(q)| \leq \epsilon N$ with probability at least $1-\delta$.
\end{corollary}

The  corollary follows directly from the theorem, by noticing that when $|S_{\cap}| < \frac{3n_{\max}\log(4pd\sqrt{B}/\delta)}{B\epsilon^2}$, then $f(q) \in \{0, \frac{4n_{\max}\log(4pd\sqrt{B}/\delta)}{B\epsilon^2}\}$ with high probability, and $\hat{f}(q) \in \{0, \frac{4n_{\max}\log(4pd\sqrt{B}/\delta)}{B\epsilon^2}\}$. Therefore, with high probability, the estimator is at most $\frac{4n_{\max}\log(4pd\sqrt{B}/\delta)}{B\epsilon^2}\}$ away of $f(q)$.

\paragraph{Complexity.}\label{par:complexity}
The space complexity of $\sketch1$ is $C = O(w \times d \times B \times b \times |\mathcal{A}|)$. Computational complexity for query execution is 
$O(p \times d \times B  \times \log(B))$, and for insertions is $O(|\mathcal{A}| \times d \times \log(B))$ -- the last logarithm is for maintaining an ordered set of samples, which speeds up insertions.

\paragraph{Configuring the sketch.}\label{sec:choosing_B}
The user sets the available memory $M$, and the values of $\epsilon$ and $\delta$. The values of $w$ and $d$ are computed as follows: $w = 1 + \lceil e * ((\epsilon + 1)/\epsilon)^{1/d}\rceil =$ $\Theta{((1/\epsilon)^{1/d})}$ and $d = \lceil \ln(2/\delta)\rceil$. In order to not exceed the memory quota $M$ (in bits), the user chooses the maximum $B$ that satisfies 
 $M \geq w * d * |\mathcal{A}| * ( 32 + B * (\lceil \log(4B^{2.5}/\delta) \rceil + 3 * 32 + 1))$. ~\footnote{To the best of our knowledge, there is no closed-form solution of this formula. However, the solution can be efficiently approximated with a numerical computation toolkit, e.g., using the bisection method.}

\subsection{Extensions}\label{sec:rangequeries}
\paragraph{Support for range queries}
All proposed sketches support range queries on numerical attributes, as well as combinations of range and point predicates, e.g., counting the number of records with $integer(112.1.4.1) \leq ipSrc \leq integer(112.1.255.255)$, \linebreak $integer(202.21.1.1) \leq ipDest \leq integer(202.22.255.255)$, and \linebreak $0 \leq dscp \leq 16$. This is achieved by indexing and querying dyadic ranges, similar to the technique used in Count-min sketches~\cite{cormode2005}. In particular, the ranges-enabled OmniSketch contains $\log(|\mathcal{D}(a_i)|)$ internal attributes sketches for each attribute $a_i$, denoted as \linebreak $CM_{(i,0)}, CM_{(i,1)}, \ldots,$  $ CM_{(i,\log(|\mathcal{D}(a_i)|)-1)}$. Each internal sketch \linebreak $CM_{(i,k)}$  keeps the frequency statistics for dyadic ranges of length $2^k$. Therefore, $CM_{(i,0)}$ stores statistics for points, $CM_{(i,1)}$ stores statistics for ranges of size 2, and so on. 

A record $r$ is indexed into the sketch as follows. For each numerical attribute $r_i$, and for $k=0$ to $\log(|\mathcal{D}(a_i)|)-1$, we find all ranges of the form $[x * 2^k+1, (x+1)2^k]$ that contain $r_i$, where $x \in \mathbb{Z}^*$. For each range of size $2^k$, we add $x$ to $CM_{(i,k)}$. Query execution follows a similar process. At query time, any range predicate is broken to its canonical cover -- all sub-ranges that follow the above form. Then, for each row of the sketch, we query all sub-ranges and merge the retrieved samples, effectively constructing a single sample that covers the complete query range. The remaining querying process remains identical to point queries.

Maintenance of the internal range sketches increases the space and time complexity of the sketch by a factor of  $\log(|\mathcal{D}(a_i)|)$. The approximation error depends on the number of dyadic ranges contained in the canonical cover of the query, which is at most $2\log(|\mathcal{D}(a_i)|)$, as in~\cite{dyadicranges}. Therefore, the total error is at most $2\epsilon\log(|\mathcal{D}(a_i)|) N$.

\paragraph{Support for general updates}
As described, OmniSketch supports the typical insert-only (i.e., cash register~\cite{muthu:fnttcs05}) data stream model. OmniSketch can be extended to handle the more general turnstile model, i.e., with updates on existing data and deletes. Incrementally maintaining the in-bucket samples then becomes the major challenge. If we have fully specified deletes (i.e., all attribute values including the rid are known at the time of deletion), then a more complex stream-sampling method like 2LHS (Section~\ref{sec:prelim}) can be employed in place of K-minwise hashing to support deletions. The case where the deletes are not fully-specified (e.g., only some attribute values are known) is more challenging to address in constant/logarithmic time, and is part of our ongoing work.

%% file: 5.experiments.tex
\section{Experimental Evaluation}\label{sec:experiments}
The purpose of our experiments was: (a) to compare the space complexity, efficiency, and accuracy of the three proposed sketches to each other and to the state-of-the-art, and, (b) to examine how our best performing sketch, $\sketch1$, performs when summarizing streams of different characteristics, and in different configurations.

\paragraph{Datasets.} For our experiments, we have used two real-world datasets (SNMP~\cite{snmp} and CAIDA~\cite{caida}), and several synthetic datasets that enabled us to thoroughly evaluate particular properties of our algorithms. SNMP contains 8.2 Million records with 11 attributes, whereas CAIDA contains 109 Million records, each with 6 attributes. The SNMP dataset contains records collected from the wireless network of Dartmouth college during the fall of 2003, whereas CAIDA is a network flow monitoring dataset, collected by an internet service provider in the USA.
We also generated synthetic datasets, with different properties/distributions, to check how the properties of the input stream affect our algorithms. Unless otherwise mentioned, the reported results correspond to the SNMP dataset. The code for dataset loading/generation/pre-processing, as well as the list of all files and attributes used in our experiments, is included in our github repository.~\footnote{\label{githubLink} https://github.com/wiegerpunter/omnisketch}

\paragraph{Queries.} We methodically constructed the queries in order to comprehensively cover a broad spectrum of query characteristics (different number of results, different number of predicates, different predicate values). At a pre-processing step, we went over a small sample of the dataset (roughly 0.05\% of all records). For each sampled record, and for each possible number of predicates $p \in [2, |\mathcal{A}|]$, we constructed 10 queries by randomly choosing $p$ attributes from the set of attributes $\mathcal{A}$, and their values in the sampled record as predicates. All queries were maintained in a set, effectively filtering out duplicates. This led to a total of 39,319 queries for the SNMP dataset on 11 distinct attributes, and 3,304 queries for CAIDA on 6 distinct attributes. More details can be found in the github repository.$^{\ref{githubLink}}$.

\paragraph{Hardware and implementation.} All experiments were executed on a linux machine, equipped with 512 GB RAM and an Intel Xeon(R) CPU E5-2697 v2, clocked at 2.7GHz. The experiments were single-threaded, i.e., only one of the 48 cores was used, and the machine was otherwise idle. All sketches were implemented and executed in Java (OpenJDK version 19.0.2). For the Hydra baseline, we used the original code of the authors~\cite{hydra}. Since Hydra was originally constructed as a Spark plugin, to avoid unnecessary extra delays imposed by Spark, we extracted and used only the code that was necessary for centralized execution. Also, to ensure a fair comparison, we excluded from Hydra's implementation any code and data structures associated with statistics beyond cardinality estimation, such as the L2 norm and entropy of sub-populations. Our changes improved Hydra's performance and reduced its space complexity. All algorithms were given a few seconds warm-up time (300 thousand updates), before we started to measure ingestion time. The code of our methods, as well as detailed instructions on processing the datasets and reproducing our results are made publicly available. $^{\ref{githubLink}}$

\subsection{Comparison of $\sketch0$ and $\sketch1$}
\label{sec:experimentsOwnSketches}

Our first series of experiments was designed for comparing the two $\sketch0$ estimators ($\sketch0_{\min}$ and $\sketch0_{\cap}$) with $\sketch1$, in terms of  (a) stream ingestion time, (b) memory requirements, (c) accuracy of the estimates, and, (d) query execution time. Accuracy was quantified by computing the mean absolute error, normalized by the stream size, i.e., $\sum_{q\in \mathbb{Q}}| \hat{f}(q) - f(q)|/(N*|\mathbb{Q}|)$, where $N$ denotes the stream size and $\mathbb{Q}$ denotes the set of executed queries. For this set of experiments, all sketches were configured with $(\epsilon=0.1, \delta=0.1)$. We tested four different configurations of $\sketch1$, with 10, 50, 100, and 200 MB of RAM.

\begin{table*}[ht]
\begin{minipage}[b]{0.49\linewidth}
\centering
    \begin{tabular}{lcc}
    \toprule
    {} &  Query  &  Observed  \\
    Algorithm            &    Time (ms)           &        Error           \\
    \midrule
    $\sketch0_{\cap}$   &     20.886 &          $4 \times 10^{-6}$ \\
    $\sketch0_{\min}$   &     29.867 &          0.0038 \\
    $\sketch1 (10 MB)$  &      0.227 &          0.0004 \\
    $\sketch1 (50 MB)$  &      1.212 &          0.0003 \\
    $\sketch1 (100 MB)$ &      2.452 &          0.0003 \\
    $\sketch1 (200 MB)$ &      6.620 &          0.0003 \\
    \bottomrule
    \end{tabular}
    \caption{Time required for query execution, and  estimation error}
    \label{tab:error_snmp}
\end{minipage}\hfill
\begin{minipage}[b]{0.49\linewidth}
\centering
    \begin{tabular}{l|llll}
\toprule
{\# query pred.} &   2   &   4   &   6  &   8 \\
\midrule
$\sketch1 (10 MB)$  & 0.18  & 0.13  & 0.12 & 0.09  \\
$\sketch1 (50 MB)$  & 0.98 & 0.79 & 0.68 & 0.44  \\
$\sketch1 (100 MB)$ & 2.02 & 1.61 & 1.25 & 1.13  \\
$\sketch1 (200 MB)$ & 6.38 & 4.33 & 2.85 & 2.64  \\
Hydra (10 MB)       & 0.04 & 0.04 & 0.02 & 0.02  \\
Hydra (50 MB)       & 0.05 & 0.04 & 0.03 & 0.00 \\
Hydra (100 MB)      & 0.05 & 0.04 & 0.02 & 0.02 \\
Hydra (200 MB)      & 0.05 & 0.04 & 0.02 & 0.00  \\
\bottomrule
\end{tabular}
    \caption{Query Execution Time in milliseconds, for queries with different numbers of  predicates $p$.}
    \label{tab:queryExecutionHydra}
\end{minipage}
\end{table*}

\begin{figure}[t]
    \centering
    \includegraphics[width=0.48\textwidth]{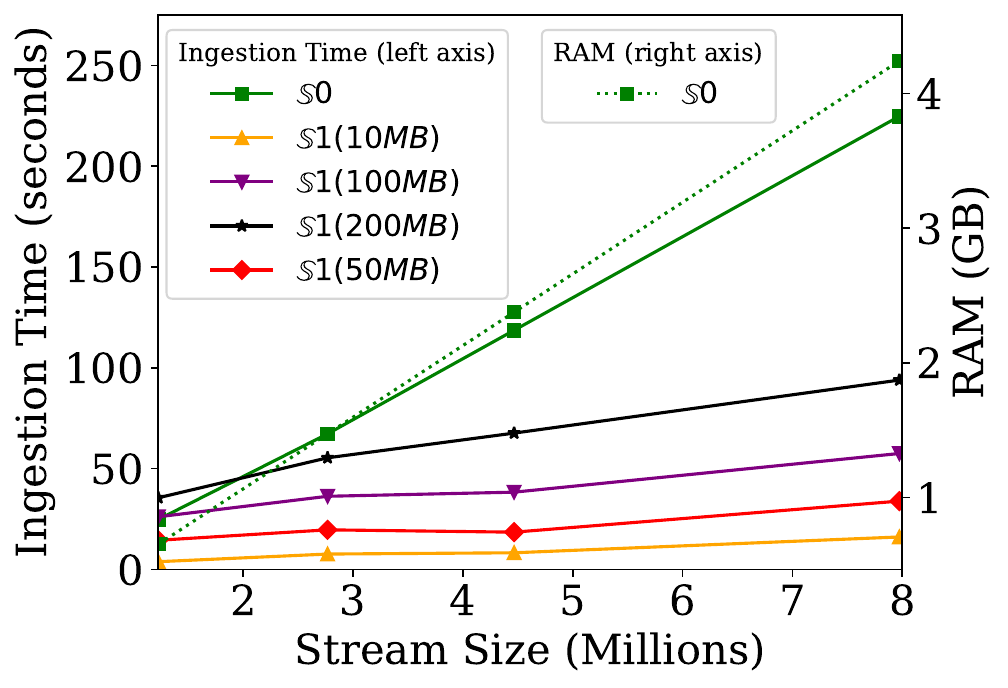}
    \captionof{figure}{Ingestion time and RAM for summarizing different stream lengths}
    \label{fig:scaling_snmp}
\end{figure}
Figure~\ref{fig:scaling_snmp} shows the required time and memory for summarizing subsets of the SNMP stream (all 11 attributes), with the compared sketches. Recall that $\sketch0_{\min}$ and $\sketch0_{\cap}$ use the same sketch, and are therefore presented together as $\sketch0$ in this figure. $\sketch1$ memory requirements are not included in the figure as separate series, since these are fixed for each configuration.  As expected, we see that the required time for building all sketches (left axis) grows linearly with the stream size. Sketch $\sketch1$ is however significantly more efficient, requiring 4 to 5 times less time compared to $\sketch0$. Also, the throughput of $\sketch1$ decreases when the sketch is allowed to use more memory. This is again expected, since a higher memory quota for $\sketch1$ translates to a higher value of $B$ (more samples per cell), which leads to slower maintenance of the red-black tree. Still, even for $\sketch1$ with 200MB RAM and 11 attributes, throughput exceeds 89 thousand updates per seconds. Also notice that $\sketch1$ ingestion time forms a subtle, yet visible, elbow (e.g., for $\sketch1$ with 200MB, this happens at around 2.7 Million updates). This elbow signifies that the K-minwise samples at most cells in the sketch reached to an almost \emph{stable} state, and the probability that any new update needs to be added to the samples is small. At this point, most updates take only $O(|\mathcal{A}|\times d)$ time, as opposed to  $O(\log(B) \times |\mathcal{A}|\times d)$.
Finally, we observe that the memory requirements of $\sketch0$ (Fig.~\ref{fig:scaling_snmp}, right axis) grow linearly with the stream size, since all records are kept in the sketch. 

Table~\ref{tab:error_snmp} summarizes the average error and query execution time for all considered configurations. We observe that both $\sketch0$ variants take significantly more time for query execution compared to the $\sketch1$ variants. The difference in performance is somewhere between 1 and 2 orders of magnitude, depending on the memory quota of $\sketch1$. The reason for this stark difference is because the $\sketch0$ variants need to iterate over very large sets for computing the intersection, whereas $\sketch1$ iterates over sample sizes of size $B$, with $B$ in the order of a few tens of thousands. We also see that $\sketch0_{\cap}$ estimator is notably faster than $\sketch0_{\min}$, even though both operate on the same sketch structure. This is attributed to the way the query execution is implemented in the two estimators. In $\sketch0_{\cap}$, similar to $\sketch1$, we start from the first cell, and keep reducing the candidate records until we intersect all $p \times d$ cells (see the relevant discussion in the implementation of $\sketch1$, Section~\ref{sec:sketch1}). This quickly reduces the candidate records. On the other hand, the implementation of the  $\sketch0_{\min}$ estimator performs the same process \emph{per row}, and takes the minimum value of all rows. Therefore,  $\sketch0_{\min}$ repeats each check multiple times, for each record that satisfies the query.

In terms of approximation error, $\sketch0_{\cap}$ outperforms all others, with a negligible average error. This is to be expected, since  $\sketch0_{\cap}$ does not suffer from underestimations and false negatives, and its probability of including false positives in the estimate is very small: $\epsilon^{p'd}$, where $p'$ denotes the number of query attributes not satisfied by the record. We also see that $\sketch1$, which relies on the intersection, has around an order of magnitude smaller error than $\sketch0_{\min}$, even though it requires one to two order of magnitudes less memory for representing the same dataset.

\paragraph{Summary.} The comparison between  $\sketch0_{\min}$,  $\sketch0_{\cap}$ and  $\sketch1$ revealed that $\sketch0_{\cap}$ substantially outperforms the other two variants in terms of accuracy, but with linear space requirements. Therefore, $\sketch0_{\cap}$ is only useful as an alternative to full indexing, where there is sufficient memory for storing the whole stream. The size of $\sketch0_{\min}$ also grows linearly with the data, and the estimator also performs worse compared to $\sketch0_{\cap}$. On the other hand, $\sketch1$ offers a favorable trade-off between accuracy and space complexity/efficiency, and allows the user to fine-tune the memory quota in order to fully utilize the available RAM. Therefore, for the remaining experiments and comparison with the state-of-the-art we will only consider $\sketch1$.

\subsection{Comparison with the state-of-the-art}
Our second series of experiments focused on comparing $\sketch1$, our best-performing sketch, with Hydra. The two sketches were allowed the same memory, and were compared on their: (a) ingestion time, (b) query execution performance, and, (c) accuracy. 

\begin{figure*}
\begin{minipage}[b]{0.49\linewidth}
    \centering
    \includegraphics[width=\textwidth]{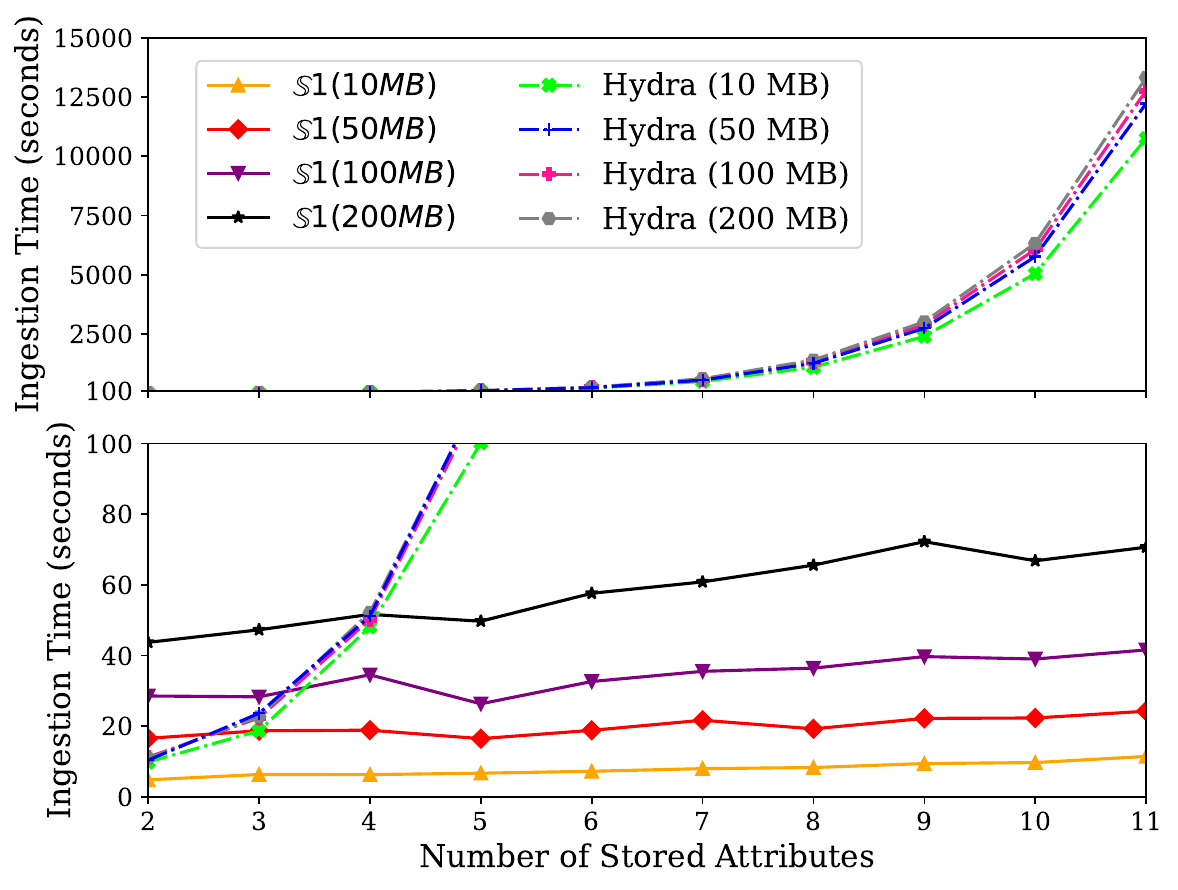}
    \caption{Ingestion Time for Hydra and $\sketch1$, for ingesting different vertical partitions (varying number of attributes) of the SNMP stream. \protect\footnotemark }
    \label{fig:ingestionTimeHydra}
\end{minipage}
\hfill
\begin{minipage}[b]{0.49\linewidth}
    \centering
    \includegraphics[width=\textwidth]{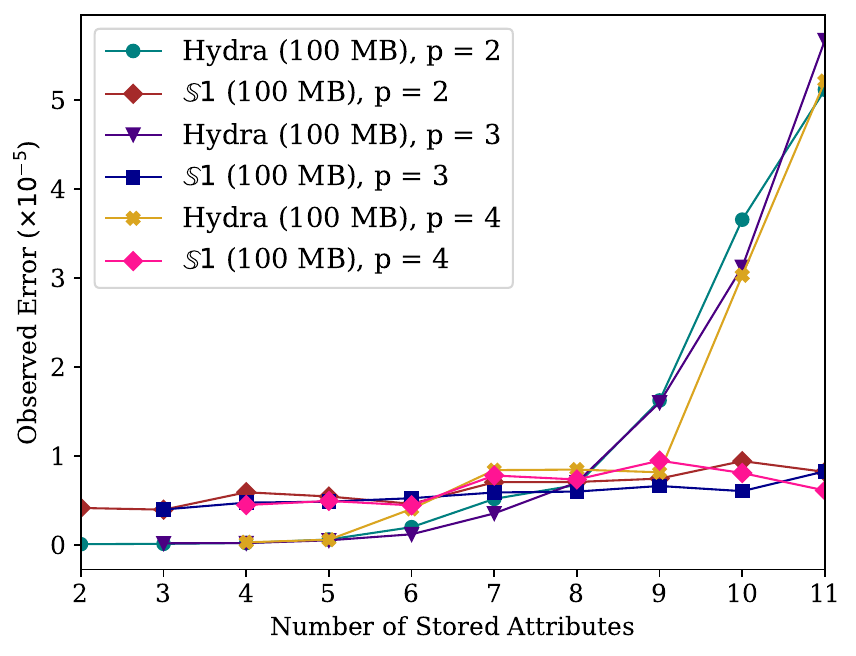}
    \caption{Estimation error of Hydra and $\sketch1$, on different vertical partitions of SNMP}
    \label{fig:hydraaccuracy}
\end{minipage}
\end{figure*}
\footnotetext{For illustration purposes, the plot is broken to two different sub-plots, with different scaling at the Y axis.}

\paragraph{Ingestion time.} 
High throughput is the key requirement for stream processing and summarization. 
Hence, our initial experiment aimed to assess the time required by $\sketch1$ and Hydra to summarize the entire stream, and to examine how this time was affected by the number of attributes in the stream. We generated streams 
with varying number of attributes (from 2 to 11) by taking distinct vertical partitions of SNMP (i.e., different subsets of the available attributes).

Figure~\ref{fig:ingestionTimeHydra} shows the ingestion time of Hydra and $\sketch1$, for  different memory quotas. For illustration purposes, the Y axis is split to two sub-ranges of different scales. Our first observation is that both algorithms have comparable performance when storing up to 4 attributes. However, the computational complexity of Hydra grows exponentially with the number of stream attributes. This happens because each record in Hydra leads to  $2^{|\mathcal{A}|}-1$ sketch updates, where $\mathcal{A}$ denotes the set of stream's attributes. Therefore, storing all 11 attributes of SNMP in Hydra leads to $O(2^{11})$ sub-populations/updates per record, and takes around 14 thousand seconds for the full dataset (more than 1 msec per record). It is important to note here that storing all sub-populations is required in Hydra, not only for supporting queries with exactly 11 predicates, but also for supporting queries that contain subsets of these predicates whenever the exact combinations of the query predicates is not known before observing the stream. 

Interestingly, addition of more attributes has a negligible impact on the ingestion time of $\sketch1$. In fact, there exist cases where adding an attribute does not increase, or even slightly reduces ingestion time (e.g., when adding the fifth attribute for $\sketch1$, 200 MB). At first sight, this result may appear counter-intuitive, since adding more attributes means maintaining more attribute sketches, and storing more samples in these sketches. This behavior is attributed to the fact that the most time-consuming step of adding a record at $\sketch1$ is the one of maintaining the K-minwise samples, which becomes faster when the sample size $B$ is reduced. By adding more attributes in the sketch without increasing its memory quota, we effectively reduce the sample size $B$ for all of the sketch's attributes. This leads to: (a) reduction of the probability that the red-black tree needs to be updated, and, (b) faster updates of the red-black tree, whenever these are required.

\paragraph{Accuracy.} 
Figure~\ref{fig:hydraaccuracy} plots the observed error on the two sketches, for queries with $p=2$, $3$, and $4$ predicates. We see that the estimation accuracy of $\sketch1$ is not significantly influenced by the number of stream attributes, whereas Hydra's accuracy suffers when the number of stream attributes increases beyond 8, even if the queries contain only 2 attributes. This behavior of Hydra is again attributed to the number of inserts that Hydra eventually performs per record: the information summarized by Hydra increases exponentially to the number of attributes, leading to more collisions and to a drastic increase of the observed error.

\paragraph{Query execution time.}
For the final comparison, we used both Hydra and $\sketch1$ to summarize all 11 attributes of the whole SNMP stream (8.3 Million updates).  Table~\ref{tab:queryExecutionHydra} summarizes the query execution time for queries with up to 8 predicates. Both sketches are very efficient, requiring  less than 10 msec for executing a query, even in the configuration with a 200MB RAM quota. We also note that $\sketch1$ is slower than Hydra. This is expected, since $\sketch1$ needs to compute the intersection of $p \times d$ samples of size $B$, whereas  Hydra only needs to compute the minimum of an approximately equal number of counters. For the same reason, the query time for $\sketch1$ also grows slightly with the available memory, whereas Hydra performance stays unaffected.  Also notice that $\sketch1$'s efficiency increases with the number of predicates, aligning with our earlier observation (Section~\ref{sec:experimentsOwnSketches}). This improvement is attributed to the algorithm's way of computing the sets intersection, which becomes more efficient as the number of attributes/sets increases (see Section~\ref{sec:sketch1}).

\paragraph{Summary.} Both Hydra and $\sketch1$ require less than 10 msec to execute queries with up to 8 predicates. However, $\sketch1$ substantially outperforms Hydra in efficiency when summarizing streams with many attributes,  with a difference in throughput that may exceed two orders of magnitude. 
Importantly, Hydra's throughput rate decreases exponentially with the number of stream attributes. The same trend of exponential decrease with the number of stream attributes is observed on Hydra's estimation accuracy, even for queries that contain very few attributes. Therefore, Hydra is not a viable option for summarizing streams that contain  many attributes. 

\subsection{Evaluation of $\sketch1$ with different streams}
Our next set of experiments focused on investigating the effect of the stream properties
(distribution of the attribute values and number of records in the stream) to $\sketch1$'s efficiency 
and accuracy.

\begin{figure*}[ht]
\begin{minipage}[b]{0.3\linewidth}
\centering
\includegraphics[width=\textwidth]{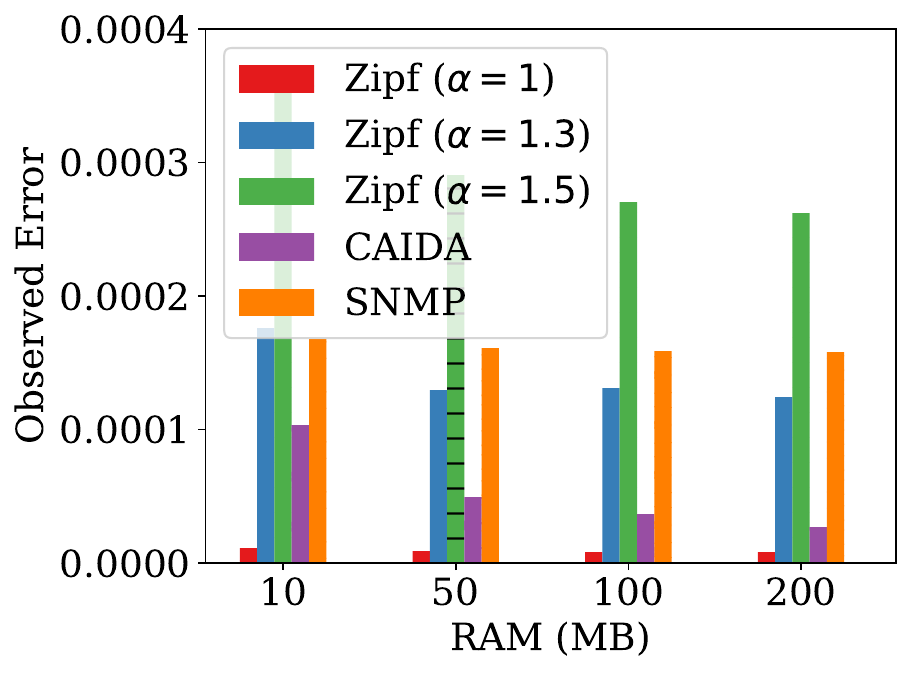}
\caption{Observed error of $\sketch1$ for  different streams.\protect\footnotemark}
\label{fig:errordatasets}
\end{minipage}\hfill
\begin{minipage}[b]{0.3\linewidth}
\centering
\includegraphics[width=\textwidth]{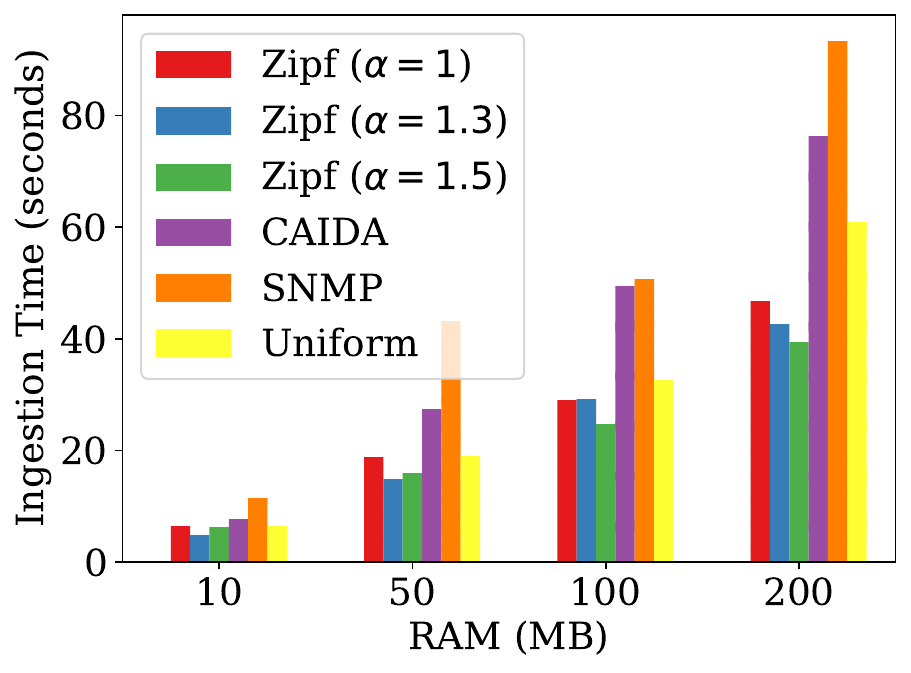}
\captionof{figure}{Ingestion time of $\sketch1$ for different streams}
\label{fig:ingestiondatasets}
\end{minipage}
\hfill
\begin{minipage}[b]{0.3\linewidth}
\centering
    \includegraphics[width=\textwidth]{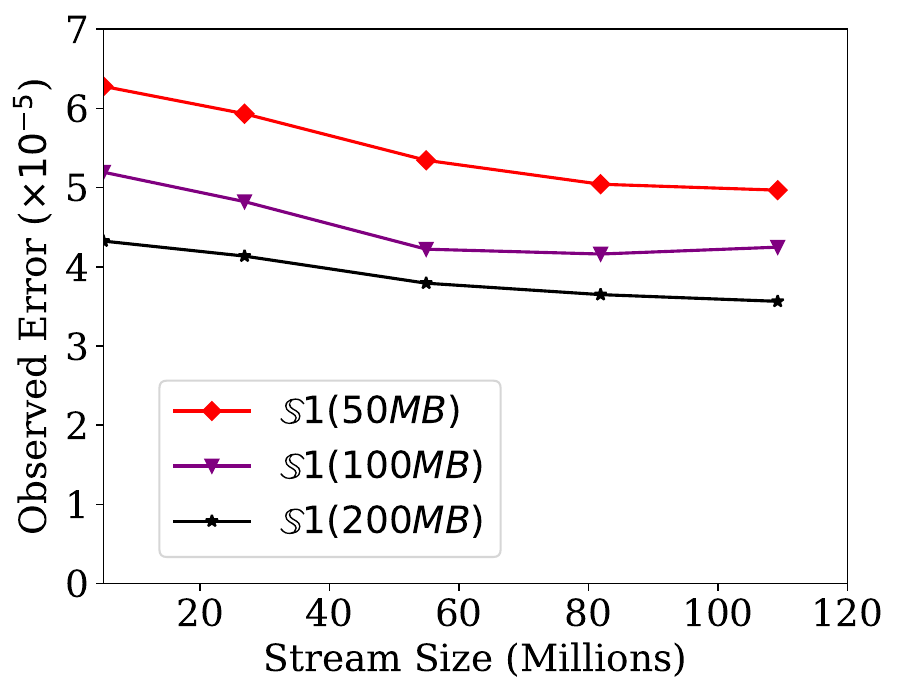}
\captionof{figure}{Estimation error of $\sketch1$ for streams of different lengths.}
\label{fig:accuracywithstreamsize}
\end{minipage}
\end{figure*}
\footnotetext{The error at the Uniform stream was close to 0 and is thus omitted from the figure.}

Figure~\ref{fig:ingestiondatasets} shows the required time for summarizing 5 Million records from the following streams:\footnote{Details for the generation of all streams are included in the project's github}
\begin{itemize}
    \item \textbf{Zipf, $\alpha \in [1, 1.3, 1.5]$}: Three different streams. Each of these streams contains 5 attributes (integer values).  We generate the records by drawing random numbers from Zipfian distributions with exponent $\alpha \in [1, 1.3, 1.5]$.
    \item \textbf{Uniform}: The stream contains 5 integer attributes. The values for all attributes are drawn by a uniform distribution.
    \item \textbf{CAIDA and SNMP}: A vertical partitioning of the CAIDA and SNMP streams, containing 5 attributes.
\end{itemize}

We see that $\sketch1$ requires less than 100 seconds to summarize each stream, even for the 200 MB quota. Also, the ingestion time for the Zipf streams is notably smaller compared to the  uniform stream, and is further reduced as the Zipf's $\alpha$ value increases.
This behavior can be traced back to the effort required for maintaining the K-minwise samples, which aligns with our earlier observation of the subtle elbow in Figure~\ref{fig:scaling_snmp}. As the Zipf exponent value rises in this experiment, the distribution of values becomes more skewed, causing more records to hash into the same cells at the corresponding attribute sketches. Consequently, the K-minwise samples of these cells swiftly reach an almost stable state. This pattern is also observed in the CAIDA and SNMP streams, where certain attributes also follow a Zipfian distribution, resulting in faster ingestion compared to the uniform stream. Still, even for the extreme case that the values for all stream attributes follow a uniform distribution, $\sketch1$  throughput surpasses 150k updates per second.

The observed  estimation error on these datasets is shown in Figure~\ref{fig:errordatasets}. The presented results correspond to the average error over 450 queries containing three predicates. We see that the observed error remains very low in all cases. For the case of Uniform, the average error is close to 0, whereas the highest error is observed with Zipf, $\alpha=1.5$. The reason for this is because, as the Zipf exponent grows, the samples of the popular cells in the sketch (the cells responsible for storing the values with frequencies at the head of the Zipf distribution) end up containing almost exclusively records of these popular values. As a result, not sufficient samples remain for low-frequency predicates that happen to fall in the same popular cells, leading to higher estimation errors. This  limitation is common across all small-memory sketches that provide error relevant to the stream size, e.g., the Count-min sketch~\cite{cormode2005}.

Our final experiment with point queries aimed to explore the impact of the stream size to the performance of $\sketch1$. Since we already demonstrated that the throughput of $\sketch1$ is unaffected with the stream size (Figure~\ref{fig:scaling_snmp}), this experiment focused solely on the impact of the stream size to accuracy. Precisely, we used $\sketch1$ to summarize CAIDA, which was the largest real dataset with  109 Million records. At regular intervals, we paused the stream ingestion, executed a fixed set of 3304 queries on the sketch, and computed the estimation error per query. The depicted results, shown in Figure~\ref{fig:accuracywithstreamsize}, correspond to the mean observed error per interval.  As predicted by the analysis, the observed error in OmniSketch stays stable with the stream size.

\paragraph{Summary.} $\sketch1$ is efficient in all cases, offering throughput that exceeds 150k updates per second. In terms of efficiency, the most difficult distribution is the uniform distribution, because it takes longer for the samples in the sketch cells to reach to a stable state. In terms of accuracy, Zipf distributions with extremely high values are more difficult for  $\sketch1$. This is a limitation shared across all small-memory sketches. Furthermore, increasing the stream size does not have a significant influence on the observed error.

\subsection{Range queries}
The final set of experiments was for evaluating the performance of $\sketch1$ for range queries.  
The queries for this experiment were generated by choosing random $p$-dimensional ranges of length 
$2^{25}$, for different values of $p$. In all cases, the starting and ending 
points of the ranges were within the minimum and maximum values of the corresponding attributes. 
In the following  we report results for $p=\{2,3\}$. 
While $\sketch1$ imposes no constraints on the values of $p$ and the range length, we noticed that  
smaller ranges or larger $p$ values led to queries with empty results. 
Also notice that not all stream attributes were suitable for range 
queries -- some attributes were categorical (e.g., attribute 
\verb+protocol+ in the IP header). Therefore, for this experiment we 
only considered the subset of attributes where range queries were meaningful.

\begin{figure}
\centering
\includegraphics[width=0.5\textwidth]{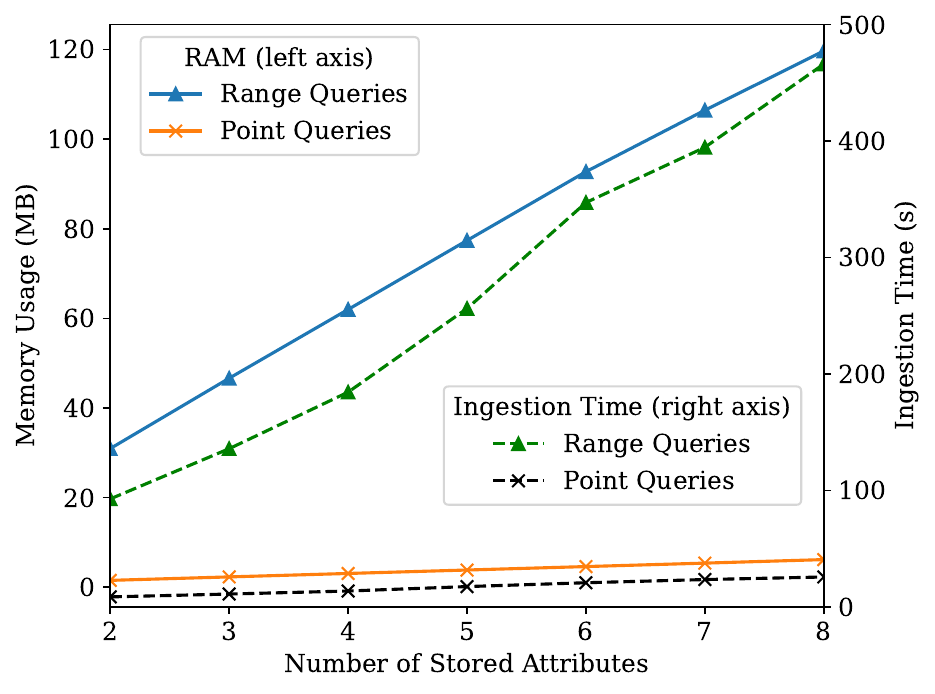}
\caption{Memory and time needed to support range queries for 2 - 8 attributes.}
\label{fig:ingestionRanges}
\end{figure}

Figure~\ref{fig:ingestionRanges} shows the required time and memory for 
summarizing the SNMP stream (attributes ifOutOctets, ifOutErrors, ifOutDiscards, ifOutUcastPkts, ifInOctets, ifInUcastPkts, ifInErrors, ifInDiscards), 
with the range-enabled $\sketch1$. The results correspond to $B=1000$, $w=20$, and $d=3$.
As an indication, the figure also includes the required time for
summarizing the same stream with the standard $\sketch1$ sketches that are not configured for 
supporting range queries.
Similar to the standard $\sketch1$, both memory and ingestion time increase linearly with the number of attributes. This is expected, since each new attribute requires a new attribute sketch. 
Also, the range-enabled $\sketch1$ is slower and requires more memory 
for keeping the same number of samples. The reason for this performance degradation is 
that in the range-enabled sketch, the number of internal attributes sketches 
is increased by a logarithmic factor, for storing statistics for the dyadic ranges. 
Interestingly, though, the observed degradation is much smaller than the
theoretical prediction. 
This discrepancy sources from the sketches responsible for storing the large dyadic ranges; 
recall that the number of large dyadic ranges is much less than the number of small 
dyadic ranges. For example, for an attribute  with domain size $2^{32}$, 
there exist only four dyadic ranges of size $2^{30}$. As such, the cells storing large dyadic ranges are mostly
empty, and require only a few bytes of RAM, whereas the few non-empty cells ($\approx 4 \times d$ per predicate) are still limited to $B$ samples. This leads to both a faster ingestion (less frequent changes at the samples) and a smaller memory footprint. 
The mean observed error for range queries with $p = 2$ is $0.094$ ($886$ queries) and for $p = 3$ is $0.072$ ($99$ queries). These are within the bound shown in Section~\ref{sec:rangequeries}.

\paragraph{Summary.} Our last set of experiments demonstrated that the range-enabled $\sketch1$ maintains high performance and high accuracy also for range queries. Storage and computational complexity of maintaining $\sketch1$ grows linearly with the number of attributes.

%% file: 6.conclusions.tex
\section{Conclusions}
\label{sec:conclusions}
We presented OmniSketch, a sketch focused on summarizing the distributions of complex streams (with many attributes) in small space. The sketch combines small (user-defined) memory footprint, and fast updating and querying times (log-linear complexity) and offers theory-backed accuracy guarantees for both point and range queries. A thorough experimental evaluation of the sketch revealed that it can achieve very fast update rates, even when summarizing streams with many attributes. For example, a sketch utilizing 200 MB can summarize an 11-attributes stream with a throughput exceeding 89 thousand updates per second, whereas the 100 MB sketch on the same stream supports twice this throughput. Furthermore, we have shown that OmniSketch outperforms the state-of-the-art (in our experiments, by more than 2 orders of magnitude in throughput) while still providing highly accurate estimates.